\documentclass[twoside]{dis08}
\usepackage[latin1]{inputenc}
\usepackage[dvips]{graphicx,epsfig,color}
\usepackage{wrapfig,rotating}
\usepackage{amssymb,amsmath,array}

\newcommand{\lsim}{\,{\buildrel < \over {_\sim}}\,}
\newcommand{\gsim}{\,{\buildrel > \over {_\sim}}\,}
\newcommand{\sqrtsNN}{\sqrt{s_{\scriptscriptstyle{{\rm NN}}}}}
\newcommand{\av}[1]{\left\langle #1 \right\rangle}

\newcommand{\mev}{\mathrm{MeV}}
\newcommand{\gev}{\mathrm{GeV}}
\newcommand{\tev}{\mathrm{TeV}}
\newcommand{\fm}{\mathrm{fm}}

\newcommand{\cm}{\mathrm{cm}}

\newcommand{\mum}{\mathrm{\mu m}}

\newcommand{\PbPb}{\mbox{Pb--Pb}}

\newcommand{\RAA}{R_{\rm AA}}

\newcommand{\pt}{p_{\rm t}}
\renewcommand{\d}{{\rm d}}
\newcommand{\dEdx}{{\rm d}E/{\rm d}x}
\newcommand{\dNdy}{{\rm d}N_{\rm ch}/{\rm d}y}

\newcommand{\ccbar}{\mbox{$\mathrm {c\overline{c}}$}}
\newcommand{\bbbar}{\mbox{$\mathrm {b\overline{b}}$}}

\newcommand{\Dz}{\mbox{$\mathrm {D^0}$}}

\newcommand{\Jpsi} {\mbox{J\kern-0.05em /\kern-0.05em$\psi$}\xspace}

\begin{document}
\title{Heavy flavour in ALICE}

\author{A. Dainese\\ (for the ALICE Collaboration)
%
%
\vspace{.3cm}\\
%
INFN - Laboratori Nazionali di Legnaro, Legnaro (Padova), Italy
}

\maketitle

\begin{abstract}
The ALICE experiment, currently in the commissioning phase, 
will study nucleus--nucleus and proton--proton collisions at the 
CERN Large Hadron Collider (LHC). We review the ALICE heavy-flavour 
physics program.
\end{abstract}

\section{Introduction}
\label{intro}

ALICE~\cite{alicePPR1,alicePPR2} is the dedicated heavy-ion experiment at the Large
Hadron Collider (LHC). The main physics goal of the experiment is the 
study of strongly-interacting matter in the conditions of high energy 
density ($>10~\gev/\fm^3$) and high temperature ($\gsim 0.2~\gev$),
expected to be reached in central \mbox{Pb--Pb} collisions at 
$\sqrtsNN=5.5~\tev$. 
Under these 
conditions, according to lattice QCD calculations, quark confinement into 
colourless hadrons should be removed and
a deconfined Quark--Gluon Plasma should be formed~\cite{alicePPR1}.

Heavy-flavour particles
are regarded as effective probes of the system conditions. 
In particular: 
\begin{itemize}
\item open charm and beauty hadrons would be sensitive to the energy density,
through the mechanism of in-medium energy loss of heavy quarks;
\item quarkonium states would be sensitive to the initial temperature of the
system through their dissociation due to colour screening.
\end{itemize}

\section{Heavy-flavour phenomenology in nucleus--nucleus collisions at the LHC}

The expected 
$\ccbar$ and $\bbbar$ production yields for pp collisions at 
$\sqrt{s}=14~\tev$ are 0.16 and 0.0072, respectively~\cite{alicePPR2}. 
For the 5\% most central Pb--Pb collisions at $\sqrtsNN=5.5~\tev$ 
the expected yields are 115 and 4.6, respectively.
These numbers, assumed as the baseline for ALICE simulation studies, 
are obtained from pQCD calculations at NLO~\cite{hvqmnr}, including 
the nuclear modification of the parton distribution functions 
in the Pb nucleus.
Note that the predicted yields have large uncertainties, of about a factor 2,
estimated by varying the values of the calculation parameters.

Experiments at the Relativistic Heavy Ion Collider (RHIC)
 have shown that the nuclear modification factor of 
particles $\pt$ distributions, 
\mbox{$R_{\rm AA}(\pt,\eta)=
{1\over \av{N_{\rm coll}}} \cdot 
{\d^2 N_{\rm AA}/\d\pt\d\eta \over 
\d^2 N_{\rm pp}/\d\pt\d\eta}$},
is a sensitive observable 
for the study of the interaction of the hard partons 
with the medium produced in nucleus--nucleus collisions.
Heavy quark medium-induced quenching is one of the most captivating 
topics to be 
addressed in \mbox{Pb--Pb} collisions at the LHC. Due to the 
QCD nature of parton energy loss, quarks are predicted to lose less
energy than gluons (that have a higher colour charge) and, in addition, 
the `dead-cone effect' is expected to reduce the energy loss of massive 
quarks with respect to light partons~\cite{dk}. 
Therefore, one should observe a pattern 
of gradually decreasing $\RAA$ suppression when going from the mostly 
gluon-originated
light-flavour hadrons ($h^\pm$ or $\pi^0$) 
to D and to B mesons: 
$\RAA^h\lsim\RAA^{\rm D}\lsim\RAA^{\rm B}$.
The measurement of D and B meson production cross sections will 
also serve as a baseline for the study of medium effects on quarkonia.
Two of the most interesting items in the quarkonia sector at the LHC 
will be: (a) understanding the 
interplay between colour-screening-induced (and medium-temperature-dependent)
suppression and statistical regeneration for 
J/$\psi$ production in a medium containing on the order of 100 $\ccbar$ pairs; 
(b) measuring for the first time 
medium effects on the bottomonia resonances, expected to be affected 
by colour-screening only if the temperature of the medium will 
be larger than about 0.4~GeV.

\section{Heavy-flavour detection in ALICE}
\label{exp}

The ALICE experimental setup, described in detail in~\cite{alicePPR1},
allows the detection of open charm and beauty hadrons and of quarkonia
in the high-multiplicity environment 
of central \PbPb~collisions at LHC energy, where a few thousand 
charged particles might be produced per unit of rapidity. 
The heavy-flavour capability of the ALICE detector is provided by:
\begin{itemize}
\item Tracking system; the Inner Tracking System (ITS), 
the Time Projection Chamber (TPC) and the Transition Radiation Detector (TRD),
embedded in a magnetic field of $0.5$~T, allow track reconstruction in 
the pseudorapidity range $-0.9<\eta<0.9$ 
with a momentum resolution better than
2\% for $\pt<20~\gev/c$ 
and a transverse impact parameter\footnote{The transverse impact parameter,
$d_0$, is defined as the distance of closest approach of the track to the 
interaction vertex, in the plane transverse to the beam direction.} 
resolution better than 
$60~\mum$ for $\pt>1~\gev/c$ 
(the two innermost layers of the ITS, $r\approx 4$ and $7~\cm$, 
are equipped with silicon pixel 
detectors).
\item Particle identification system; charged hadrons are separated via 
$\dEdx$ in the TPC and via time-of-flight measurement in the 
Time Of Flight (TOF) detector; electrons are separated from charged 
hadrons in the dedicated
Transition Radiation Detector (TRD), and in the TPC; 
muons are identified in the muon 
spectrometer covering the pseudo-rapidity range $-4<\eta<-2.5$. 
\end{itemize}

Simulation studies~\cite{alicePPR2}
have shown that ALICE has good potential to carry out
a rich heavy-flavour physics programme. The main analyses in preparation 
are:
\begin{itemize}
\item Open charm (section~\ref{open}): fully reconstructed hadronic decays 
$\rm D^0 \to K^-\pi^+$, $\rm D^+ \to K^-\pi^+\pi^+$,
$\rm D_s^+ \to K^-K^+\pi^+$ (under study), $\rm \Lambda_c^+ \to p K^-\pi^+$ (under study) in $|\eta|<0.9$.
\item Open beauty (section~\ref{open}): 
inclusive single leptons ${\rm B\to e}+X$ 
in $|\eta|<0.9$ and ${\rm B\to\mu}+X$ in $-4<\eta<-2.5$; inclusive displaced
charmonia ${\rm B\to J/\psi\,(\to e^+e^-)}+X$ (under study).
\item Quarkonia (section~\ref{quarkonia}): $\rm c\overline c$ (J/$\psi$, 
$\psi^\prime$) and $\rm b\overline b$ ($\Upsilon$, 
$\Upsilon^\prime$, $\Upsilon^{\prime\prime}$) states 
in the ${\rm e^+e^-}$ ($|\eta|<0.9$) and $\mu^+\mu^-$ ($-4<\eta<-2.5$) 
channels.
\end{itemize} 
For all simulation studies, a multiplicity $\dNdy=4000$--$6000$
was assumed for central \mbox{Pb--Pb} collisions.
In the following, we report the results corresponding to the 
expected statistics collected by ALICE per LHC year: 
$10^7$ central (0--5\% $\sigma^{\rm inel}$) \mbox{Pb--Pb} events at
$\mathcal{L}_{\rm Pb-Pb}=10^{27}~\cm^{-2}{\rm s}^{-1}$
and $10^9$ pp events at 
$\mathcal{L}_{\rm pp}^{\rm ALICE}=5\times 10^{30}~\cm^{-2}{\rm s}^{-1}$,
in the barrel detectors; the forward muon arm will collect
about 40 times larger samples of muon-trigger events
(i.e.\, $4\times 10^8$ central \mbox{Pb--Pb} events).

\section{Charm and beauty measurements}
\label{open}

\begin{figure}[!t]
  \begin{center}
    \includegraphics[width=0.52\textwidth]{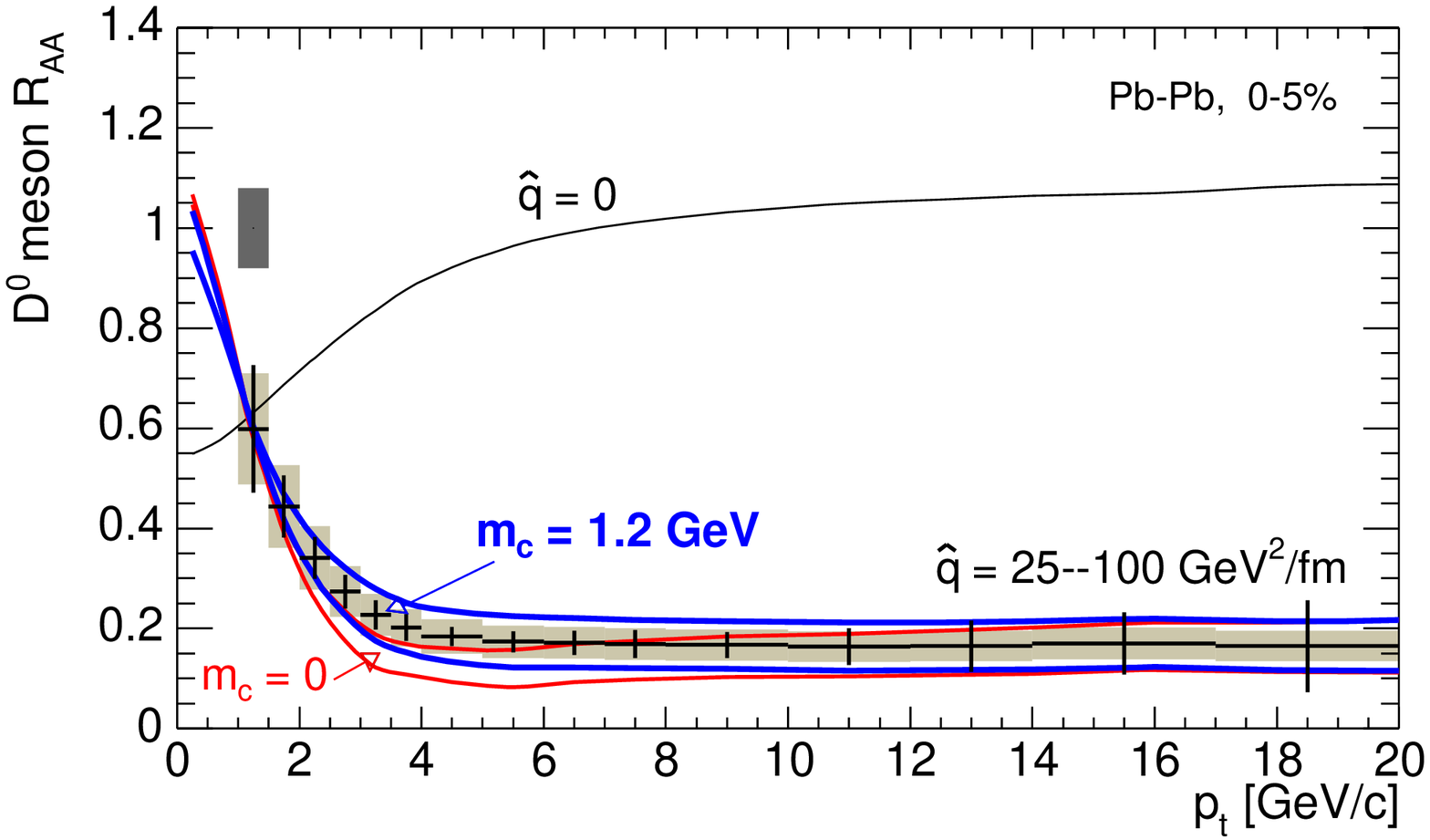}
    \includegraphics[width=0.47\textwidth]{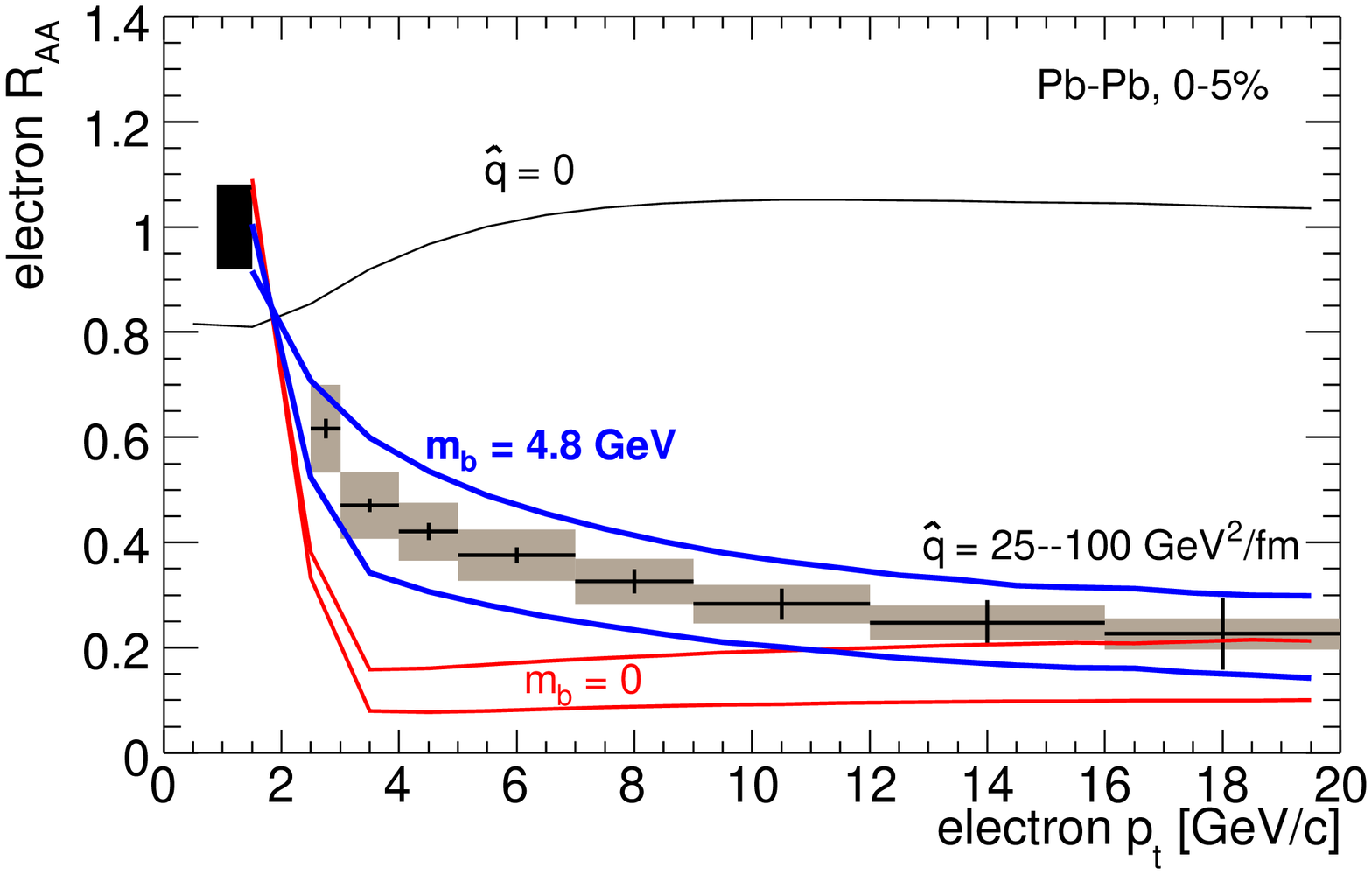}
    \caption{Nuclear modification factors for $\Dz$ mesons 
             (left) and for B-decay electrons
             (right).
             Errors corresponding to the centre of the prediction bands   
             for massive quarks are shown: bars = statistical, 
             shaded area = systematic.} 
    \label{fig:RAA}
  \end{center}
\end{figure}

Among the most promising channels for open charm detection are the 
$\rm D^0 \to K^-\pi^+$ ($c\tau\approx 120~\mum$, branching ratio 
$\approx 3.8\%$) and $\rm D^+ \to K^-\pi^+\pi^+$ ($c\tau\approx 300~\mum$, 
branching ratio $\approx 9.2\%$) decays. The detection strategy
to cope with the large combinatorial background from the underlying event 
is based on the selection of displaced-vertex topologies, i.e. separation 
from the primary vertex of
the tracks from the secondary vertex 
and good alignment between the reconstructed D meson momentum 
and flight-line~\cite{alicePPR2,elena}. 
An invariant-mass analysis is used to extract the raw signal 
yield, to be then corrected for selection and reconstruction efficiency
and for detector acceptance.
The accessible $\pt$ range for the $\Dz$ is $1$--$20~\gev/c$ in \mbox{Pb--Pb} and 
$0.5$--$20~\gev/c$ in pp, 
with statistical errors better than 15--20\% at high $\pt$. Similar capability 
is expected for the $\rm D^+$.
The systematic errors 
(acceptance and efficiency corrections, 
centrality selection for Pb--Pb) are expected to be smaller than 20\%.

The production of open beauty can be studied by detecting the 
semi-electronic decays of beauty hadrons, mostly B mesons. 
Such decays have a branching ratio of $\simeq 10\%$.
The main sources of background electrons are: decays of D mesons; 
$\pi^0$ Dalitz decays 
and decays of light vector mesons (e.g.\,$\rho$ and $\omega$);
conversions of photons in the beam pipe or in the inner detector 
layer; pions misidentified as electrons. 
Given that electrons from beauty have average 
impact parameter $d_0\simeq 500~\mum$
and a hard $\pt$ spectrum, it is possible to 
obtain a high-purity sample with a strategy that relies on:
electron identification with a combined $\dEdx$ (TPC) and transition
radiation (TRD) selection;
impact parameter cut to 
reduce the charm-decay component and 
reject misidentified $\pi^\pm$ and $\rm e^{\pm}$
from Dalitz decays and $\gamma$ conversions.
As an example, with $10^7$ central 
\mbox{Pb--Pb} events, this strategy is expected to allow
the measurement of electron-level 
$\pt$-dif\-fe\-ren\-tial cross section in the range $2<\pt<20~\gev/c$ 
with statistical errors smaller
than 15\% at high $\pt$. Similar performance figures are expected for 
pp collisions.

B production in pp and \mbox{Pb--Pb} collisions 
can be measured also in the ALICE muon 
spectrometer ($-4<\eta<-2.5$) analyzing the single-muon $\pt$ 
distribution~\cite{alicePPR2}.
The main backgrounds to the `beauty muon' signal are $\pi^\pm$, 
$\rm K^\pm$ and charm decays. The cut $\pt>1.5~\gev/c$ is applied to all
reconstructed muons in order to increase the signal-to-background ratio.
Then, a fit technique allows to extract a $\pt$ distribution of muons 
from B decays.
Since only minimal cuts are applied, the statistical errors are 
expected to be smaller than 5\% up to muon 
$\pt\approx 30~\gev/c$.

We investigated the possibility of using 
the described charm and beauty measurements 
to study the high-$\pt$ suppression induced by parton energy loss.
The sensitivity to $\RAA^{\rm D}$ 
and $\RAA^{\rm e~from~B}$ is presented in Fig.~\ref{fig:RAA}.
Predictions~\cite{adsw} with and without the effect of the
heavy-quark mass, for a medium transport coefficient $\hat{q}$ 
(a measurement of the medium density) in the range 
$25$--$100~\gev^2/\fm$, are also shown.

\section{Quarkonia capabilities}
\label{quarkonia}

ALICE can detect quarkonia 
in the dielectron channel at central rapidity ($|y|\lsim 1$) and in the
dimuon channel at forward rapidity ($-4\lsim y\lsim -2.5$). In both channels
the quarkonia 
acceptance extends down to zero transverse momentum, since the 
minimum $\pt$ for e and $\mu$ identification is about $1~\gev/c$. 
The high $\pt$ reach 
is expected to be 10~(20)~$\gev/c$ for the J/$\psi$ in 
$\rm e^+e^-$ ($\mu^+\mu^-$), for a \mbox{Pb--Pb} 
run of one month at nominal luminosity.
In the bottomonium sector, the mass resolution of about $90~\mev$ 
at $M_{\ell^+\ell^-}\sim 10~\gev$, for both dielectrons and dimuons, 
should allow the separation of the 
$\Upsilon$ and $\Upsilon^\prime$ states, and thus the measurement of 
the $\Upsilon^\prime/\Upsilon$ ratio, which is expected to be sensitive
to the initial temperature of the medium.

Simulation studies are in progress to prepare 
a measurement of the fraction of J/$\psi$
that feed-down from B decays. Such measurement can be performed 
by studying the separation of the dilepton pairs in the J/$\psi$ 
invariant mass region
from the main interaction vertex. The analysis is also expected to provide a 
measurement of the beauty $\pt$-differential cross section.

\section{Summary}

Heavy quarks, abundantly produced at LHC energies, 
will allow to address several issues at the heart of 
in heavy-ion physics. 
They provide tools to probe the density
(via parton energy loss and its predicted mass dependence)
and the temperature
(via the dissociation patterns of quarkonia)
of the high-density QCD medium formed in \mbox{Pb--Pb} collisions.
The excellent tracking, vertexing and particle identification performance 
of ALICE will allow to fully explore this rich phenomenology.

\begin{footnotesize}

\end{footnotesize}



\begin{thebibliography}{0}

\bibitem{alicePPR1}
  ALICE Collaboration, Physics Performance Report Vol.~I,  
  CERN/LHCC 2003-049 and J.~Phys.~{\bf G30} 1517  (2003).

\bibitem{alicePPR2}
  ALICE Collaboration, Physics Performance Report Vol.~II,
  CERN/LHCC 2005-030
  and J.~Phys.~{\bf G32} 1295 (2006).

\bibitem{hvqmnr} 
   M.L. Mangano, P. Nason and G. Ridolfi, 
   Nucl. Phys. {\bf B373} 295 (1992).

\bibitem{dk}
  Yu.L. Dokshitzer and D.E. Kharzeev, Phys.~Lett. {\bf B519} 199 (2001).

\bibitem{adsw}
  N. Armesto, {\it et al.}, Phys.~Rev.~{\bf D71} 054027 (2005).

\bibitem{elena}
  E. Bruna, Int. J. Mod. Phys. {\bf E16} 2097 (2007).


\end{thebibliography}
\end{document}